\documentclass[12pt]{iopart}
\usepackage{graphicx}

\def \ee{\end{equation}}
\def \be{\begin{equation}}

\begin{document}

\title[]{Black Holes at LHC?}

\author{Ben Koch$^{1,2}$, Marcus Bleicher$^1$ and Horst St\"{o}cker$^{1,2}$}
{\normalsize
\hspace*{-8pt}$^1$ FIAS--Frankfurt Institute for Advanced Studies, Max von Laue-Str. 1,\\
D--60438 Frankfurt am Main, Germany\\
\hspace*{-8pt}$^2$ Institut f\"{u}r Theoretische Physik, Johann Wolfgang Goethe -
Universit\"{a}t,\\
D--60438 Frankfurt am Main, Germany\\[0.2ex]
}

\begin{abstract}
Strategies for identifying speculative
mini black hole events (due to large extra dimensions) 
at future colliders are reviewed.
Estimates for production cross sections, Hawking radiation,
di-jet suppression and multi- mono-jet emission are surveyed.
We further report on a class of effective entropy formulas that
could lead to the formation of a final black hole
remnant state, BHR. Such BHRs could be
both electrically charged and uncharged.
Charged BHRs should be observable by single stiff charged tracks
in the detectors. 
Collinear hadronic jets with a large missing
transverse momentum are presented as new
observable signal for electrically neutral black holes.

\end{abstract}

\section{Introduction}\label{intro}

Black holes have received
an ever growing attention, since their first prediction
from the Schwarzschild solution  \cite{Schwarzschild}.
Just recently it was conjectured that 
in the presence of  additional
compactified large extra dimensions (LXDs) \cite{add}
black holes (BH)
might even be produced in future colliders \cite{Banks:1999gd,Giddings3,ehm,dim}
like the Large Hadron Collider (LHC).
Measuring black hole physics in the laboratory
would give a unique key to test our understanding
of Planck-scale physics and quantum gravity.
Here we review the status of this field of research in
view of its possible impact on the first year of p-p running at
the LHC, including the implications of the possible existence
of black hole remnants \cite{39,40,Barrow,my2,arg,Callan,Rizzo:2005fz,axionic,Bonanno:2006eu,Hossenfelder:2001dn,Hossenfelder:2003dy,own2,Hossenfelder:2005bd,Hst06,Hst062,Alberghi:2006qr,Betz:2006ds,Hossenfelder:2005ku,Koch:2005ks,Humanic:2006xg}.

\section{Hierarchy-problem and large extra dimensions}
There exist several models which go beyond the Standard Model (SM)
by assuming extra spatial dimensions \cite{add,rs1}.
These models provide a solution
to the so-called hierarchy problem by the statement that
the huge Planck-scale, as derived from Newton's gravitational
constant $G_N$, is due to the higher dimensional geometry of space-time,
and therefore just an "effective mirror" of the true fundamental scale ($M_f$)
of gravity. This fundamental scale might be as low as a few TeV.
Although, these models can partly be motivated by String
Theory \cite{Antoniadis:1990ew},
a phenomenologist's point of view, of studying 
effective theories of some unknown deeper theory
(not necessarily String Theory)
is also justified.

In the further discussion, the model suggested 
by Arkani-Hamed, Dimopoulos and Dvali
\cite{add} is used.
In this model the $d$ extra space-like dimensions
are compactified on tori with radii $R$.
Gravitons are allowed to propagate freely in the (3+d)+1-dimensional bulk
while all SM 
particles are confined to our 3+1-dimensional sub-manifold (brane). 
The  fundamental mass $M_f$ and the  Planck mass $m_{Pl}$ are then related by
\begin{eqnarray}
m_{Pl}^2 = M_f^{d+2} R^d \quad. \label{Master}
\end{eqnarray}
This equation allows to estimate the radius $R$ of these extra dimensions.
For two extra dimensions and $M_f \sim$~TeV,
 $R$ can be as large as $2$~mm. For a constant $M_f$ a higher number of
extra dimensions corresponds to a smaller compactification radius $R$.
For recent 
updates on parameter constraints on $d$ and $M_f$ see e.g. Ref.\ \cite{Cheung:2004ab}.

\section{Black holes at the large hadron collider?}

\begin{figure}[htb]
\begin{minipage}{7.5cm}
\includegraphics[width=7.5truecm]{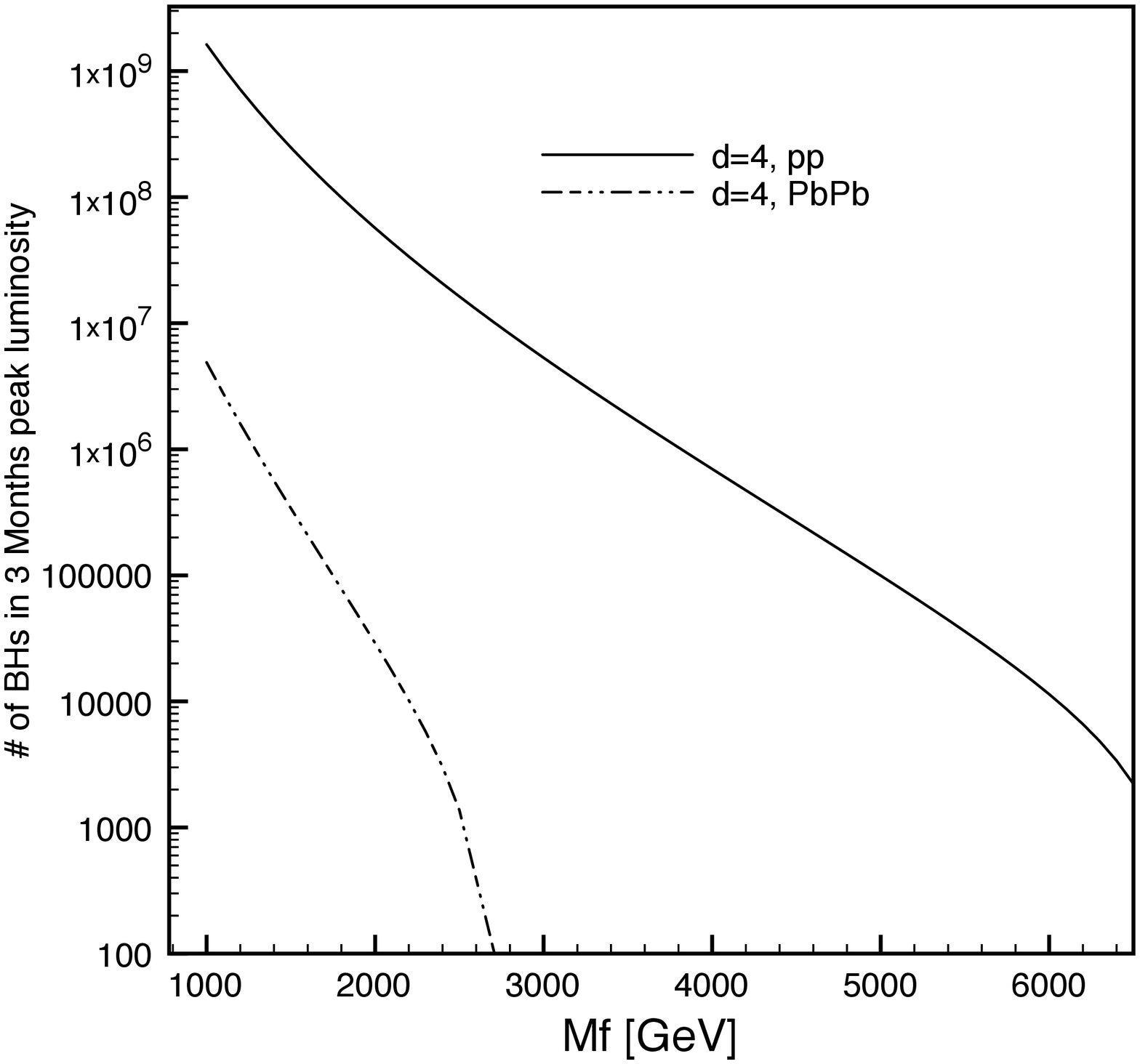}
\end{minipage}
\begin{minipage}{7cm}
\includegraphics[width=7.5truecm]{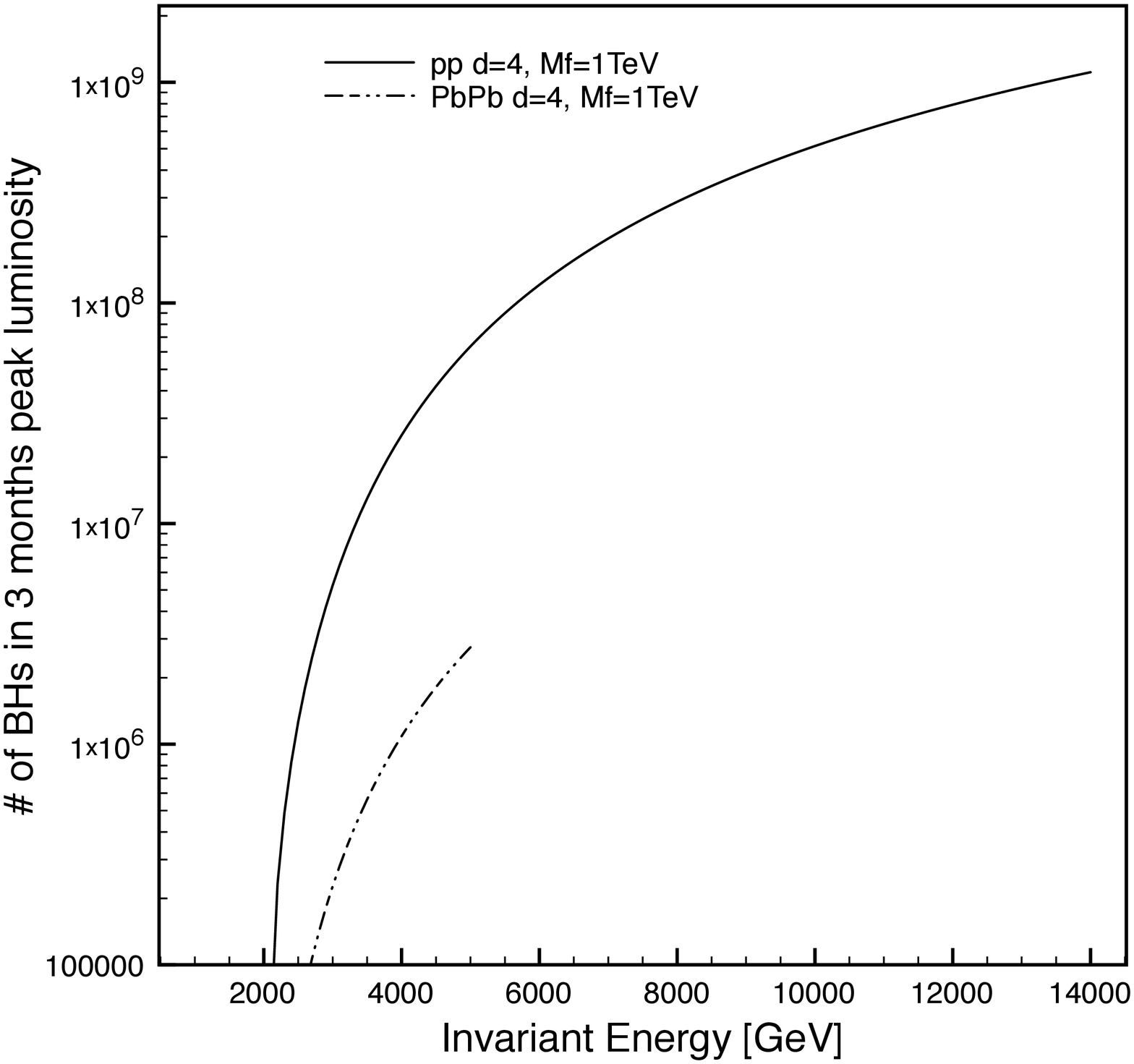}
\end{minipage}
\caption{
The left plot shows the number of black holes produced in three
months at the
LHC as a function
of the varying mass scale $M_f$.
This is done under the assumption of  a pp (PbPb) peak luminosity of
$10^{34}$ $\rm{cm}^{-1}\rm{s}^{-1}$ ($0.5 \cdot10^{27}$ $\rm{cm}^{-1}\rm{s}^{-1}$) at an invariant energy of
$14$~ATeV ($5$~ATeV).
The right plot shows the number of black holes produced in three months
under the assumption of  a pp (PbPb) peak luminosity of
$10^{34}$ $\rm{cm}^{-1}\rm{s}^{-1}$ ($0.5\cdot10^{27}$ $\rm{cm}^{-1}\rm{s}^{-1}$) 
as a function of the invariant energy $\sqrt{s}$.
In both cases, the curves for various
$d$ differ from the above depicted ones by less than a factor 10.\label{dsdm}}
\end{figure}

At distances smaller than the size of the extra
dimensions the Schwarzschild radius \cite{my} is given by
\begin{equation} \label{ssradD}
R_H^{d+1}=
\frac{2}{d+1}\left(\frac{1}{M_{f}}\right)^{d+1} \; \frac{M}{M_{f}}
\quad .
\end{equation}
This radius is 
much larger than the Schwarzschild radius corresponding  
to the same BH mass in 3+1 dimensions. 
From the Hoop conjecture \cite{hoop}
one assumes the formation of a black hole as soon
the impact parameter of two colliding particles is smaller
than the corresponding Schwarzschild radius. 
Accordingly, this minimal impact parameter
rises  enormously in the extra-dimensional setup.
The straight forward approximation of the 
LXD-black hole production cross section 
can be made by taking the classical geometric 
cross section
\begin{eqnarray} \label{cross}
\sigma(M)\approx \pi R_H^2 \quad,
\end{eqnarray}
which only contains the fundamental scale as a coupling constant.
Although, this classical cross section has been under debate
\cite{Voloshin:2001fe,Rychkov:2004sf,Jevicki:2002fq}, semi-classical considerations, 
which take into account that only a fraction of the
initial energy can be captured behind the Schwarzschild-horizon,
yield form factors of order one 
\cite{Formfactors}.
Also 
angular momentum 
considerations change the results only by a factors 
order one \cite{Solo} and the naive classical result 
remains even valid in String Theory \cite{Polchi}.
The differential cross section in proton-proton collisions is then given by 
summation over all possible parton interactions and
integration over the momentum fractions, where the kinematic relation $x_1 x_2 s=\hat{s}=M^2$ has
to be fulfilled. This yields
\begin{eqnarray} \label{partcross}
\frac{{\rm d}\sigma}{{\rm d}M}  
&=&  \sum_{A_1, B_2} \int_{0}^{1} {\rm d} x_1 \frac{2 \sqrt{\hat{s}}}{x_1s} f_A(x_1,\hat{s}) 
f_B (x_2,\hat{s})  \sigma(M,d)   \quad.
\end{eqnarray}
The particle distribution functions for $f_A$ and $f_B$ are tabulated e.g. in the CTEQ-tables \cite{cteq}.
A numerical evaluation of this expression results in the differential
cross section as shown in Fig. \ref{dsdm} (left). Most of the black holes
have masses close to the production threshold. This is due to the fact that even in high energetic
collisions, the proton contains a high number of small $x$ gluons which dominate the
scattering process. 
It is now straightforward to compute the total cross section and
the production rate by integration over Eq. (\ref{partcross}), see 
Fig. \ref{dsdm}.

%
Thus, cross sections like Eq.~(\ref{cross})
lead to the exciting prediction 
that if large extra dimensions do actually exist,
a large number (up to $10^9$ per year) of black holes will be produced in
future colliders
\cite{Banks:1999gd,dim,Hossenfelder:2001dn,Hossenfelder:2003dy,own2,Hossenfelder:2005bd,Hst06,Hst062,Alberghi:2006qr,Betz:2006ds,Hofmann:2001pz,Hossenfelder:2003jz,Casadio:2001dc,Alexeyev:2002tg,BleicherNeu,Chamblin:2002ad,Casanova:2005id,Casadio:2001wh,Hossenfelder:2004ze,Stojkovic:2004hp} 
and should in fact, be daily produced
in ultra high energetic cosmic ray events \cite{cosmicrayskk,cosmicraysbh}.

\section{Black hole evaporation process}

As exciting the sole production of microscopical black holes in
collider experiments is, to relate it to experiment
the evaporation process of the black hole is unknown.
This evaporation process is often classified in three phases.
In the balding phase, the newly formed black hole
radiates away its angular momentum by gravitational radiation.
The second phase is the Hawking phase, 
where it is assumed to enter the semi-classical regime of
quantum theory on the background of curved space-time.
According to the Hawking law,
the black hole emits radiation that is distributed by
the thermal spectrum
\begin{eqnarray}\label{eq_eomm}
\varepsilon = \frac{\Omega_{(3)}}{(2 \pi)^3}
\int  \frac{\omega^3 ~{\mathrm d}\omega }{{\rm exp}[S(M)-S(M-\omega)] + s}  \quad,
\end{eqnarray}
where  $\omega^3/({\rm exp}[S(M)-S(M-\omega)] + s)=n(\omega, \;M)$
is the spectral density and $s=0$ (or $s=\pm1$) is the Maxwell-Boltzmann, 
Fermi-Dirac, Bose-Einstein factor. 
This is true up to grey body factors up to the order of one \cite{Ida:2006tf}.
As long the BH mass $M$ is much bigger than the fundamental mass
$M_f$ the difference in the BH entropies $S(M)$ can be
approximated by a derivative leading to the Hawking temperature $T_H$:
$S(M) - S(M-\omega) \approx
(\partial S)/(\partial M) \omega =  (\omega)/(T_H)$. 
%
As soon as the BH mass approaches the fundamental mass $M\approx M_f$,
this temperature would exceed the actual mass of the black hole.
This reflects the fact that the BH entered the regime of quantum gravity,
in which no predictive theory is known and the BH's behavior and fate is unclear,
so we rely on the rough, intuitive estimates of such speculative scenarios.
In a numerical approach on could either assume
that the
BH performs a prompt final decay into $2-6$ particles which carry the BH's charge,
momentum and other quantum numbers \cite{Harris:2003db,Cavaglia:2006uk},
or that a remnant (BHR) is left over.\\
The idea of a remnant has been put forward
to cure the information loss problem.
This remnant idea is supported by
arguments employing the uncertainty relation \cite{39,40,Barrow},
by introducing
corrections to the BH-Lagrangian \cite{Callan,Rizzo:2005fz}, 
by the consideration of axionic charge \cite{axionic}, by leading
order quantum gravity considerations \cite{Bonanno:2006eu},
or by quantum hair \cite{hair}
arguments.
These arguments are mostly made for $3+1$ dimensions, but
also apply for cases where $d> 0$.
A rough intuitive modeling of the formation
of a black hole remnant with a mass $M_{R}\ge M_f$
can be done by imposing the condition 
\begin{equation}\label{condi}
M-\omega \ge M_{R}\quad,
\end{equation} 
on any single particle
emission \cite{Hossenfelder:2005ku,Koch:2005ks}. 
The spectral density with such a condition is plotted as the dotted line 
in Fig. \ref{ptexamp} left.
It is also possible to soften the rough condition 
(\ref{condi}) and to impose that the
spectral density smoothly approaches the remnant mass $M_R$.
Therefore it has to fulfill 
\be\label{eq1}
\lim_{\omega\rightarrow M-M_R}n(\omega,M)=0\quad
\mbox{and}\;
\lim_{\omega\rightarrow M-M_R}\partial_{\omega}(n(\omega,M)\omega^3)=\hbox{finite}\quad.
\ee
If one  demands that for $M\gg M_{R}$ Hawking's result is recovered, one finds
for $s=0$ that the entropy can be expressed
in terms of a Laurent series:
\be\label{eqSsol}
S(M)=\frac{d+1}{d+2}\left(\frac{M-M_R}{M_f}\right)^{\frac{2+d}{1+d}}
+\frac{1}{M_f}\int\limits_{\epsilon}^{M-M_R+\epsilon}
\sum\limits_{n=0}^{\infty}a_n \left(\frac{M_f}{x}\right)^{n+1} dx
\quad,
\ee
where $\epsilon$ is an infinitesimal positive number and $a_i$
are the coefficients of the Laurent series.
In Fig. \ref{ptexamp} left, the spectral densities for several of those
cases with non zero
coefficients $a_0$ and $a_1$, are plotted. 
The  spectral densities may allow for a more realistic
simulation of the decay of a microscopic BH into a stable BHR.

\section{Signatures for black hole formation at the LHC 
through di-jet suppression and production of multiple mono-jets}

\begin{figure}[htb]
\begin{minipage}{7.5cm}
\includegraphics[width=7.5truecm]{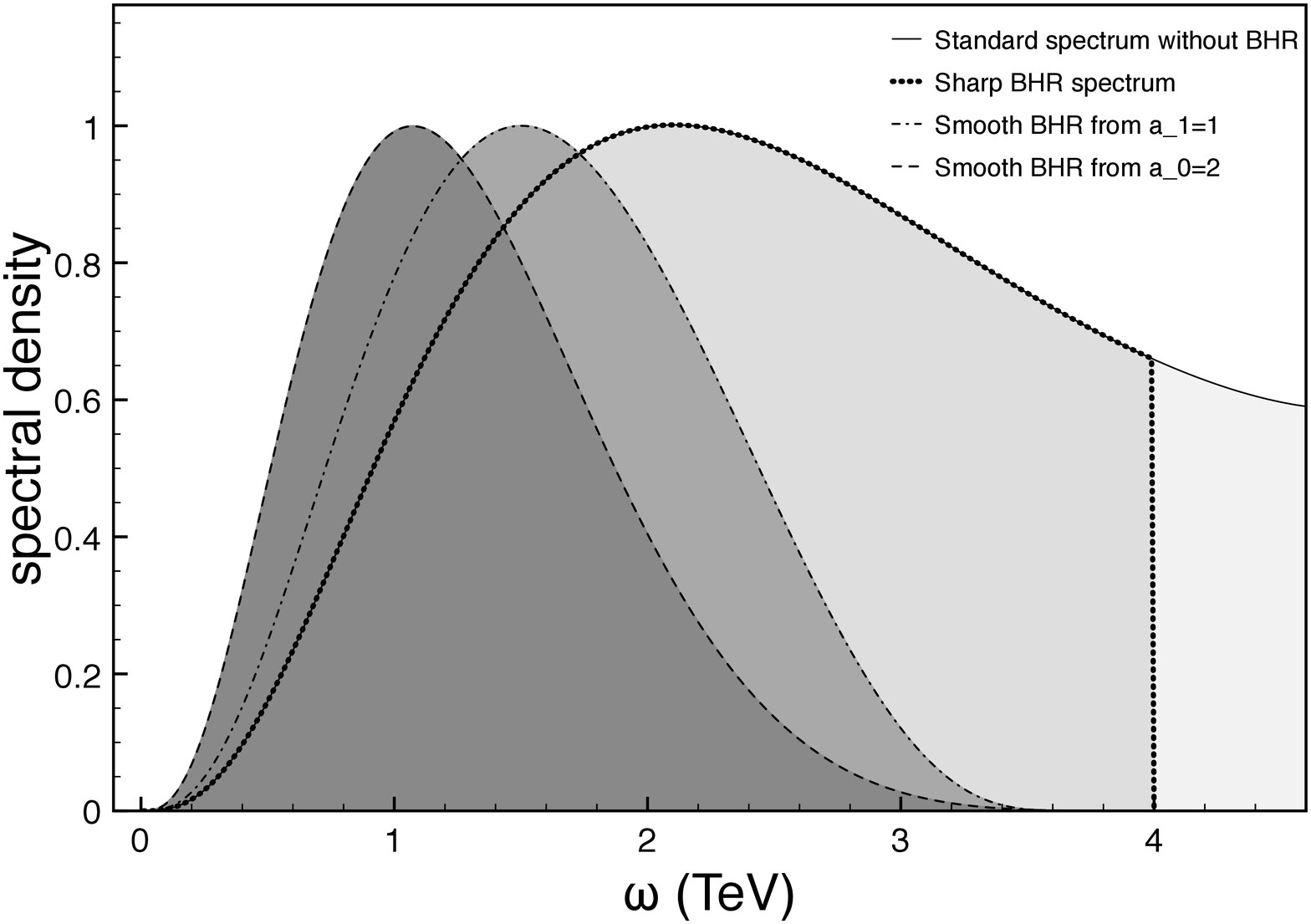}
\end{minipage}
\begin{minipage}{7cm}
\includegraphics[width=7.5truecm]{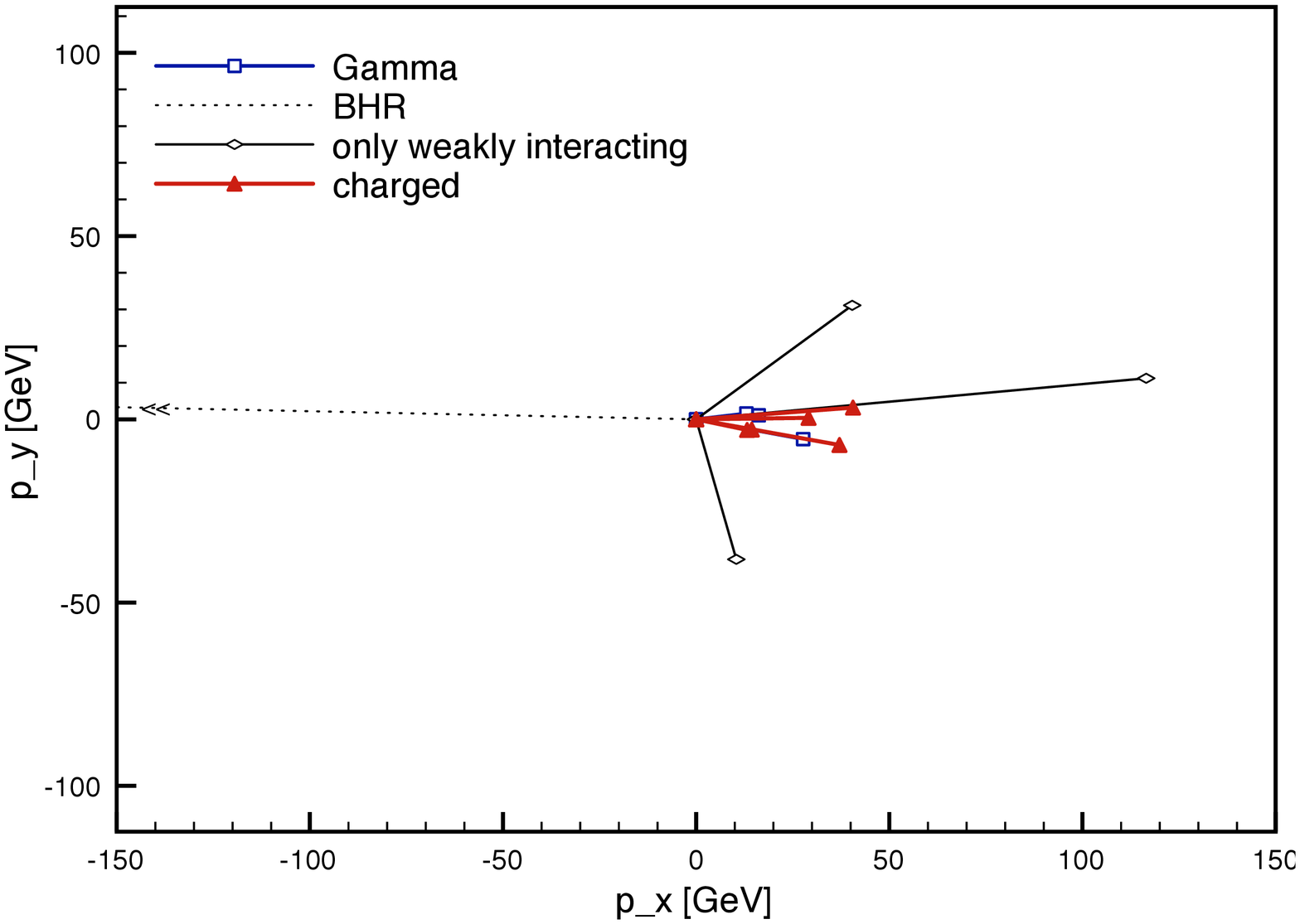}
\end{minipage}
\caption{
Left figure:
Normalized spectral densities from 
Eq. (\ref{eq_eomm}) with $M_R=4$TeV, the condition
(\ref{condi}) gives the dotted line, the modified entropy (\ref{eqSsol}) with $a_0=2$
(denoted by a\_0)
gives the dashed line, and  $a_1=1$ gives the dashed-dotted line, where all other coefficients $a_i=0$.
Right figure:
 Transverse momentum distribution of a single BH event at
the LHC with an initial 
energy of $2$~TeV and a BHR mass of $1$~TeV and $p_T>10$GeV.
The dashed line represents the BHR transverse momentum which,
in the case of a neutral BHR,
would be not visible in the detector \cite{kochPHD}.
\label{ptexamp}}
\end{figure}

\begin{figure}[htb]
\includegraphics[width=8truecm]{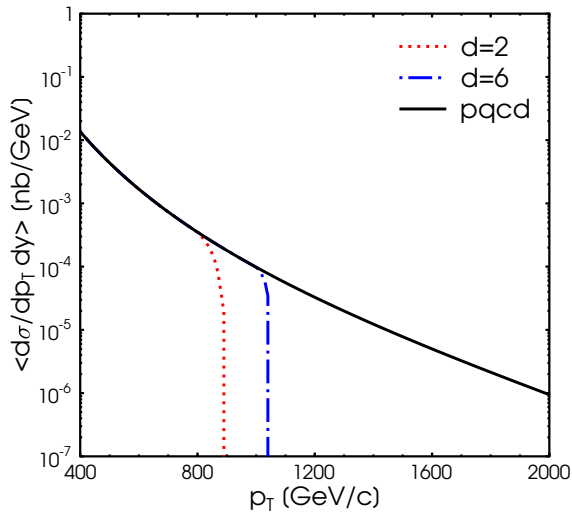}
\caption{
Differential cross section for the production of hard di-jets
with high transverse momenta. The dashed line is for two extra dimensions
and the dashed-dotted line is for six extra dimensions,
both for $M_f=1$TeV \cite{Hossenfelder:2004af}.
\label{JetSuppr}}
\end{figure}

One of the first signatures of BH formation
suggested was the suppression of hadronic
di-jet events above the BH production threshold 
(at $2 E_T>M_f$) energy
 \cite{own2,Hofmann:2001pz,Casadio:2001wh},
as can be seen in Fig. \ref{JetSuppr}. 
Here, the expected standard model cross section for jet production is 
 shown as a full line for pp interactions at $\sqrt s=14$~TeV. 
The dashed line and the dashed-dotted line 
depict the expected cut-off behavior if black hole formation is included \cite{own2}. 
The Hawking radiation from the decay of the black hole will be emitted predominantly 
around transverse momenta of $\sim 50-500$~GeV and can therefore not mask the high $p_T$ cut-off
from the black hole formation.
However, 
particles originating from the Hawking radiation
in the $p_T$ range below the
$\sim 1$~TeV cut-off
should cause multiple mono-jets (see e.g. Fig. \ref{ptexamp}, right).
\section{Signatures for black hole remnants}

The formation of stable BHRs would 
provide interesting new signatures
that allow for the identification of such a BHR event at future colliders:\\
Electrically charged BHRs would leave a stiff ionizing track in the
detector. This would allow to identify
the BHR \cite{Humanic:2006xg} and measure it's mass directly.\\
Neutral BHRs
could be identified e.g. by the
$p_t$ distributions,
multiplicities, and angular correlations \cite{Koch:2005ks,Humanic:2006xg} of the Hawking evaporated 
SM particles.
Here we propose a new signal for uncharged BHRs, namely the
search for events with $\sim$TeV missing energy plus
a quenched high $p_T$ hadron spectrum in the same event.
Here the BHR carries a major fraction
of the total energy. While many extensions of the
standard model predict missing energy signatures, here
the spray of awayside hadronic Hawking-jets, above
a 10 GeV $p_T$ cut off, shows a clear focussing, see Fig \ref{ptexamp}, right.
Such events constitute, according to our simulation
a significant fraction of the BHR events.
%
As most BHs are expected to be produced close the
the production threshold, $M_{BH}\sim M_f$,
the total event structure would then be dominated by
this particular BHR event topology.
%
%

\section{Summary}
We have surveyed observable consequences of recent
speculations of abundant black hole formation, as a
consequence of large extra dimensions, which were suggested
as a possible solution of the hierarchy problem.
Such a scenario, suppression of hard (TeV)
di-jets above the BH formation threshold should be observed at the
LHC. Most BHs are expected to be produced close to
the production threshold.
Rare high mass BHs ought to decay rapidly by multiple
hard mono jets, due to Hawking radiation.
Speculations about the formation of BH remnants
can be tested experimentally at the LHC:
Charged stable BHRs would leave single
stiff tracks in the LHC detectors, e.g. ALICE, ATLAS, and CMS.
Uncharged BHRs with their very small reaction cross sections
could be observed by searching for events with $\sim 1$~TeV missing
energy and quenching of the high $p_T$ hadron spectra.\\
Naturally, to date the dynamics of the quantum gravitational process of the BH-formation
is far from being understood.
Recently, it has been suggested that the bremsstrahlung
due to collapse of galactic, and microscopic black holes
might be so strong that BHs are not formed at all in a finite
time in the frame of a distant observer \cite{Vachaspati:2007hr}. 
Possibly, for the microscopic black hole with large extra dimensions
to be studied at the LHC this could lead to strong non equilibrium
radiation into the forward - backward direction.
This non thermal quantum radiation
might be similar to the thermal radiation with an effective,
angular dependent temperature.
\\ \\
This work was supported by GSI and BMBF.\\

\end{document}